\documentclass[aps,prstab,twocolumn,superscriptaddress]{revtex4-1}
\usepackage{amssymb}
\usepackage{amssymb}
\usepackage{amssymb}
\usepackage{graphicx}
\usepackage{graphics}
\usepackage{amsmath}
\usepackage{placeins}
\usepackage{hyperref}
\usepackage[dvipsnames]{xcolor}
\usepackage{wasysym}

\begin{document}

\title{Field emission microscopy of carbon nanotube fibers: evaluating and interpreting spatial emission}
	\author{\firstname{Taha Y.} \surname{Posos}}
	\email{posostah@msu.edu}
	\affiliation{Department of Electrical and Computer Engineering, Michigan State University, East Lansing, MI 48824, USA}
		\author{\firstname{Steven B.} \surname{Fairchild}}
	\affiliation{Materials and Manufacturing Directorate, Air Force Research Laboratory, WPAFB, OH 45433, USA}
   	\author{\firstname{Jeongho} \surname{Park}}
	\affiliation{Materials and Manufacturing Directorate, Air Force Research Laboratory, WPAFB, OH 45433, USA}
	\author{\firstname{Sergey V.} \surname{Baryshev}}
	\email{serbar@msu.edu}
	\affiliation{Department of Electrical and Computer Engineering, Michigan State University, East Lansing, MI 48824, USA}

\begin{abstract}
In this work, we quantify field emission properties of cathodes made from carbon nanotube (CNT) fibers. The cathodes were arranged in different configurations to determine the effect of cathode geometry on the emission properties.  Various geometries were investigated including: 1) flat cut fiber tip, 2) folded fiber, 3) looped fiber and 4) and fibers wound around a cylinder. We employ a custom field emission microscope to quantify I-V characteristics in combination with laterally-resolved field-dependent electron emission area. Additionally we look at the very early emission stages, first when a CNT fiber is turned on for the first time which is then followed by multiple ramp-up/down. Upon the first turn on, all fibers demonstrated limited and discrete emission area. During ramping runs, all CNT fibers underwent multiple (minor and/or major) breakdowns which improved emission properties in that turn-on field decreased, field enhancement factor and emission area both increased. It is proposed that breakdowns are responsible for removing initially undesirable emission sites caused by stray fibers higher than average.  This initial breakdown process gives way to a larger emission area that is created when the CNT fiber sub components unfold and align with the electric field. Our results form the basis for careful evaluation of CNT fiber cathodes for dc or low frequency pulsed power systems in which large uniform area emission is required, or for narrow beam high frequency applications in which high brightness is a must.
\end{abstract}

\maketitle

\section{Introduction}
New and novel cathodes are being investigated for use as electron beam sources for next generation vacuum electronic devices (VEDs).  Applications such as electron microscopy, X-ray sources, and traveling wave tube amplifiers require high current, high brightness electron beams with a narrow energy distribution.  Cathodes need to be robust and durable to protect against damage from ion back-bombardment and heating (external or self-induced) during operation. Cathode lifetimes of a few 1000’s of hours are required\cite{1}.

As VEDs progress towards higher frequency and higher power operation the benefits of using field emission cathodes rather than thermionic cathodes becomes apparent.  This primarily stems from the fact that cathode size scales as as 1/$f$, where $f$ is fundamental operating frequency of the device.  Higher frequency devices\cite{2,3,4} therefore require smaller cathodes, and the excessive heat generated by thermionic  can result in severe thermal stress placed on cathode assemblies  which leads to beam instability.  Field emission cathodes also offer the potential of fast ON/OFF switching, as compared to externally heated thermionic sources which require a temperature thermal ramp-up to reach maximum emission current.  This fast ON/OFF switching capability offers the potential of more efficient gating techniques.

Fibers made from carbon nanotubes (CNT) have demonstrated significant potential for use as field emission cathodes\cite{1,5}. CNT fibers have excellent electrical and thermal conductivity,  and produce high output emission currents with good current stability for ultralow turn-on voltage.  To date, most data on the emission properties of CNT fibers has been obtained by  measuring emission current in a simple diode configuration with the voltage applied to a metallic anode positioned above a vertically mounted fiber. However, there are numerous examples demonstrating that field emission is often not laterally uniform\cite{6,7}. Thus, there is a need to evaluate emission area to realistically estimate current density and cathode brightness. 

To spatially resolve the emission properties of CNT fibers we utilize a projection type field emission microscopy apparatus that can both measure and image the emission current.. Four different CNT fiber cathode designs were fabricated for this measurement.  We observe that CNT fibers undergo a conditioning process that immensely improve formal emission characteristics (turn-on field and field enhancement $\beta$-factor) but not necessarily spatial uniformity/coherence of emission. We find that the field emission area is responsible for unconventional emitter behavior, namely, emitter saturation and self-heating. Results and conclusions are consistent across all the tested geometries. The most promising CNT fiber cathode design is emphasized.

\section{Samples and Experimental}

\begin{figure*}
\includegraphics[scale=0.45]{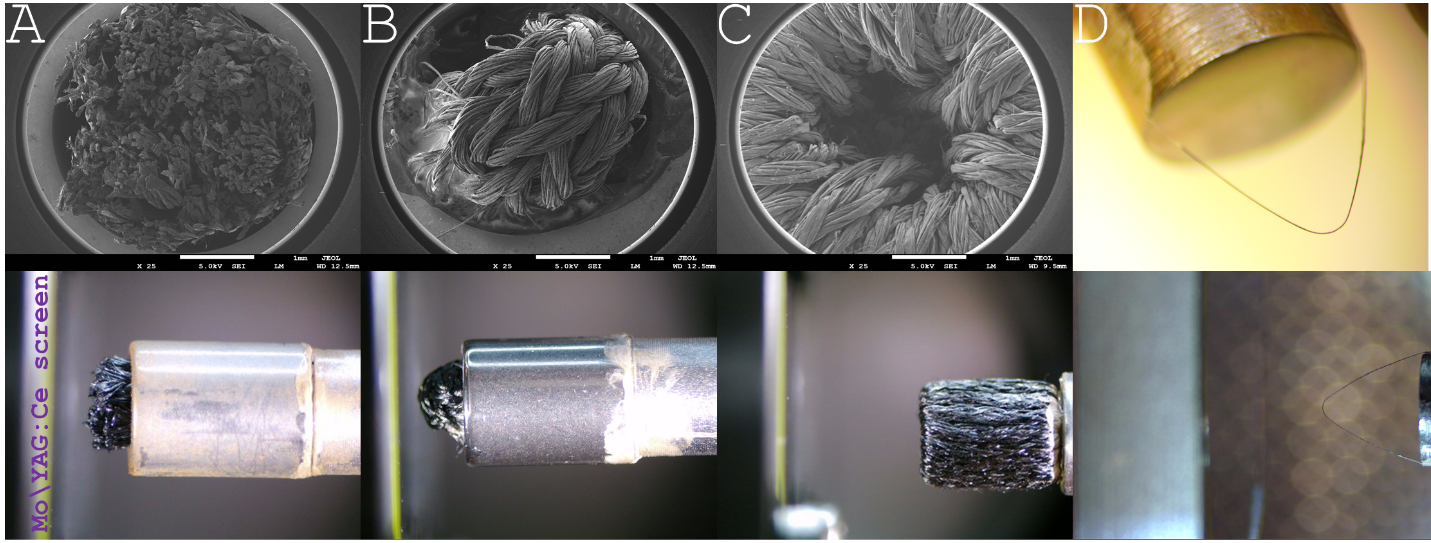}
\caption{\label{f2}SEM images of sample A (flat cut sample); sample B (folded sample); sample C (the wound geometry sample); and optical microscope image of sample D (looped sample). Bottom row: side camera views of the samples placed against the imaging YAG screen. All these images are taken before starting the experiments. There were no visible signs of unfolded stray fibrils on the samples.}
\end{figure*}

The CNT fibers used in these experiments were purchased from DexMat, Inc. in Houston, TX. The fibers were fabricated using a wet spinning technique described by Behabtu et al.\cite{8}. This fabrication process ensures that the CNTs comprising the fibers are closely packed and highly aligned which ensures high electrical and thermal conductivity\cite{25} as well as optimal performance when used as eiher wire conductors or field emission cathodes\cite{24,8,9,10}. Carbon nanotube yarns are made by twisting or braiding together multiple CNT fibers. Both individual fibers and twisted yarns were used in these experiments.

The CNTs fibers were arranged in four different configurations which utilized either a single $\sim$90$\mu$m diameter fiber or multiple fibers braided together into a larger diameter yarn.  These different configurations allowed us to investigate the effects of surface geometry on electron brightness, beam size, emission area and current density.

Sample A consisted of four yarns inserted together into a 3 mm diameter metal tube. Each yarn consisted of $\sim$300 fibers braided together to make the total yarn diameter $\sim$900 $\mu$m. The yarns are protruding from the end of the tube where then mechanically cut in an attempt to get a surface with uniform emitter heights. This was difficult to achieve due to the toughness of the CNT yarns which makes them difficult to cut. The final results are shown in Fig.\ref{f2} which shows an SEM image of the cut fibers as well as an optical image which shows a side view of the cathode.

Sample B was made of CNT yarns that were $\sim$200 $\mu$m in diameter. They consisted of 21 CNT fibers braided together. Several yarns were folded together and then shoved through the cylinder to make a somewhat rounded tip that protruded through the end of the cylinder. Fig.\ref{f2} shows an SEM image of the bunched yarns at the top of the cylinder as well as an optical image which shows a side view of the cathode.

Sample C was made by winding a CNT yarn around the wall of a 3 mm diameter metal cylinder. The purpose of this sample was to see if we could make a uniform emission edge around the edge of the cylinder. Fig.\ref{f2} shows an SEM image of the top of the cylinder showing the yarns pulled over the edge. Also shown is an optical image of the side of the cathode. Samples A,B, and C were all attached to the steel cylinder with silver paint to ensure an electrical contact to ground.

Sample D was a single looped CNT fiber of $\sim$90 $\mu$m in diameter which was arched and attached from both ends to the stainless steel base. The fiber was contacted to the steel base with silver paint. Optical images of the cathode are shown in Fig.\ref{f2}.

\begin{figure*}
\includegraphics[scale=0.5]{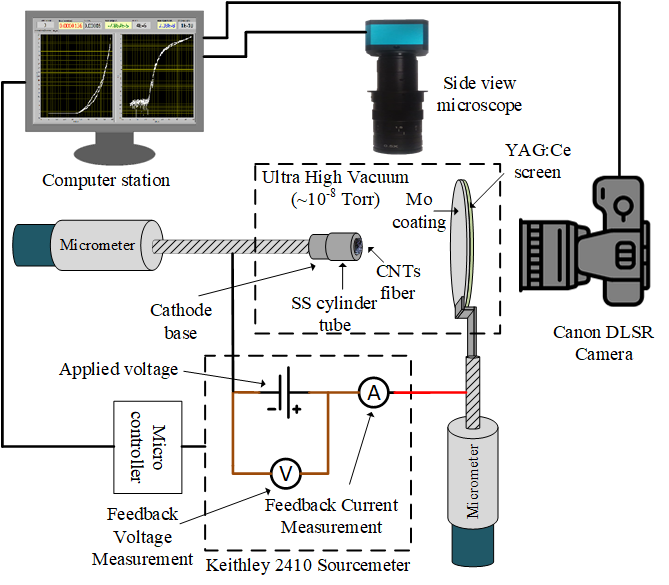}
\caption{\label{f1}Experimental setup cartoon}
\end{figure*}

The experiments were performed using a field electron microscopy technique given in Ref.\cite{11}. The measurement setup is shown in Fig.\ref{f1}. In place of a standard metal anode, we used a custom scintillator anode screen. The screen is yttrium aluminum garnet doped with cerium (YAG:Ce) coated with Molybdenum (Mo). YAG:Ce has diameter of 1 inch, and thickness of 100 $\mu$m. The Mo coating was applied in house using magnetron sputtering in UHV base pressure system, and resulting film had thickness of 7-8 nm. Metal needs to be deposited on YAG:Ce screen to make it conductive, then electric field can be establish between anode and cathode to accelerate electrons through ultra-high vacuum, and capture and them into the ground. The coating is thin enough to let electrons penetrate through Mo to YAG:Ge to produce green light and thick enough to prevent YAG:Ce screen from charging up\cite{11}. Mo is chosen because it has high melting point (2896 K), so it can sustain in high power density electron beam. No visible electron bombardment induced damage (burn through pinholes) was observed on the screen upon completing measurements. Cathode base, which samples are mounted on top of, was surface polished 316 stainless steel cylinder and 4.4 mm in diameter. The cathode base was then attached to a in-vacuum micrometer used to set the interelectrode gap. Parallelism of the screen and the sample surfaces is checked by top and side view cameras when installing the cathode (see Fig.\ref{f2}, bottom row). Samples and vacuum chamber are grounded. The screens are positioned using another translation arm that is attached to the system using a custom quartz nipple and therefore electrically isolated from the chamber. It is positively biased in the experiment. Emitted electrons from the sample under effect of bias voltage are accelerated toward the screen and strike the screen with an energy equal to the applied voltage. In such way, electrons arriving from different points of the emitting cathode surface originated at different angles create cathodoluminescence patterns (at 550 nm luminescence line) on the YAG:Ce screen. The patterns, captured by a Canon DLSR camera with CMOS full frame sensor installed at viewpoint behind the screen, represent laterally resolved field electron emission. Applied voltage, feedback current and feedback voltage readings are enabled by Keithley 2410 electrometer. In all experiments, the electrometer was programmed to sweep voltage up/down with 1 V step with 100 $\mu$A set as an upper limit for the emission current. Dwell duration for each voltage step is 5 s to sample and record current, set and feedback voltage, vacuum pressure and calculate statistical error bars. The system was programmed to take field emission images every 10V step, such that taking images was synchronized with the electrometer. All measurements were done in vacuum (2 to 5)$\times10^{-8}$ Torr.

\section{Results and Discussion}

\subsection{“Conditioning” micro-breakdowns}

\begin{figure*}
\includegraphics[scale=0.99]{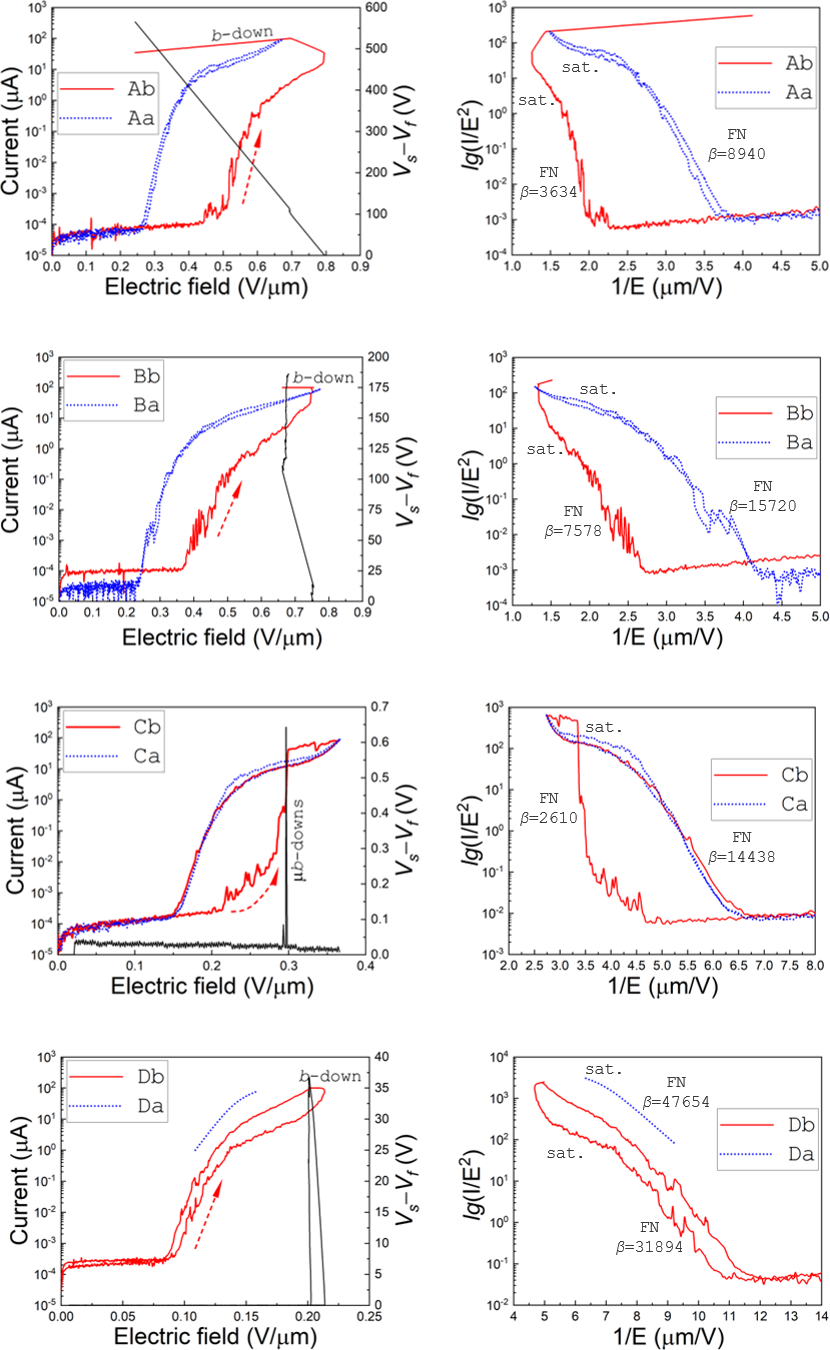}
\caption{\label{f3}Semi-log $I-E$ curves and FN plots for the studied fibers.}
\end{figure*}

All samples were tested multiple times; each test included the voltage sweep up and then down. Fig.\ref{f3} summarizes electric $I-E$ characteristics comparing the first and the last run; E-field is the actual field that is calculated using the measured feedback voltage $V_{f}$ and the measured gap. One particular feature can be seen – it is the improved efficiency of the cathodes in that the turn-on field decreased and field enhancement $\beta$-factor increased. The main vehicle mechanism of the improvement is the “conditioning” process that happens through a series of igniting/quenching emitters that, in most extreme cases, is accompanied by breakdowns of different strength. The ignition/quench process appears as extensive noise of the $I-E$ curves of the initial run for all samples, labeled as Ab, Bb, Cb, Db where b stands for “before”. Aa, Ba, Ca, Da where a stands for “after” show $I-E$ curves upon completing 4 runs. The extensive conditioning noise is visible because every point on the $I-E$ curve is collected for 5 seconds to gain enough statistics in order to calculate average current, voltage, pressure and their error bars\cite{11}: such a long dwell time captures ups and downs in the output current of the fibers turned on for the first time. The current noise of a relatively large amplitude (3 to 5 times) is associated with breakdowns (if any) that have negligible strength, i.e. cannot be detected in our system. Following our previous work\cite{12}, the sizable micro-breakdown/discharge taking place can be visualized by plotting the difference between the set voltage $V_{s}$ and the feedback voltage $V_{f}$ versus the feedback voltage or the actual E-field. Such a plot traces the voltage loss in the system due to arcing: since the electrometer is power limited, the arc will cause $V_{f}$ to drop with respect to $V_{s}$. Note, since the dwell or integration time per point is 5 s, shorter surges will result in smaller delta between $V_{f}$ and $V_{s}$ even if the breakdown/arc/discharge strength was of the same magnitude. In that sense, we are looking for non-zero difference between $V_{f}$ and $V_{s}$ to mark off the breakdown rather than evaluate its actual strength. In Fig.\ref{f3}, the $I-E$ curves are superimposed with $V_{s}$-$V_{f}$ traces. As can be seen, all four samples underwent through breakdowns of different strengths or lengths (or both). Upon the first turn on, the samples A and B do not have ramp down curves as the strength/lasting of the breakdowns was extensive and the power supply was automatically shut down via a safety interlock. Even though the breakdown is often seen a damaging process, in the present case there were significant emission property improvements. For example, the sample A before (Ab) and after (Aa) experienced 2-fold decrease of the turn-on field, from about 0.5 to 0.25 V/$\mu$m, and 2-fold increase of the $\beta$-factor, from about 3,000 to 9,000.

\subsection{Field emission microscopy and conditioning}

\begin{figure*}
\includegraphics[scale=0.45]{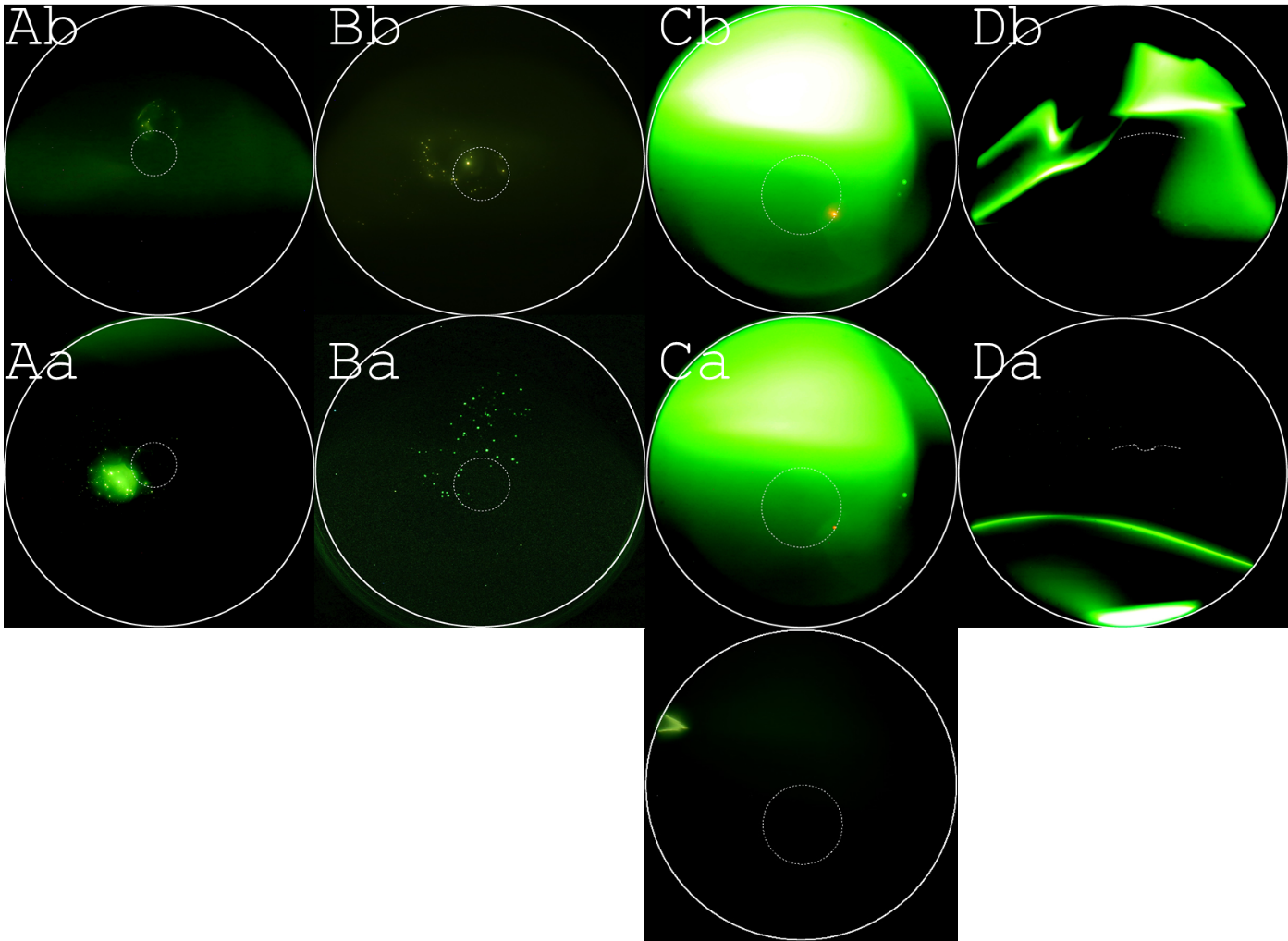}
\caption{\label{f4}The laterally resolved field emission pattern on Mo$\backslash$YAG:Ce screen taken at the same electric field before and after conditioning for sample A (0.67 V/$\mu$m), B (0.72 V/$\mu$m), C (0.36 V/$\mu$m) and D (0.16 V/$\mu$m). The white dashed circles and line show actual position and orientation of the samples with respect to the YAG screen. The outstanding image at the bottom illustrates the source of the halo background: it is a stray emitter pair projected to be nearly parallel to the screen plane thus generating electron rays that have long path across the screen resulting in intense halo.}
\end{figure*}

To better understand the effects of conditioning and fully characterize the fiber design, the presented $I-E$ curves are compared to the laterally resolved field emission micrographs that are compiled in Fig.\ref{f4}. They compare the emission patters between the first and the last tests. A few main features can be noted as follows:

1) Samples A and B improved their emission by means of increasing the total number of strong emitters seen as bright spots on the micrographs Ab/Aa and Bb/Ba. The larger the number of strong emitters (higher $\beta$-factor) the lower the turn-on field: the electrometer senses currents above the detection threshold and therefore larger number of high $\beta$-factor emitters will deliver an output current of a magnitude above the threshold at a lower E-field.

2) Sample C behaved differently. As seen from Fig.\ref{f4}, the run Cb demonstrated very slow response to the field in that the output current remained $\sim$1-10 nA even though the applied field significantly changed (corresponding to the applied voltage of 100 V, out of entire sweep ranging 0 to 375 V). Then a series of micro-breakdowns took place (at least two were detected) and the output current instantaneously inflated by over 3 orders of magnitude. Concurrently with the breakdown at 0.3 V/$\mu$m one strong emitter (see Cb in Fig.\ref{f4}) appeared as a red spot. Our imaging screens are semi-transparent to the red, and in this case red light emission from the emitting locations was bright to the extent that the green light emission from the YAG anode screen was not seen. The intense red light emission suggested that this specific emitter was delivering major portion of the detected output current $\sim$100 $\mu$A. The small emitter size (single nanotube or a cluster of single CNTs) resulted in extensive current density and therefore led to exceptional thermal heating of this emitter. Unlike samples A and B, sample C retained a very similar emission area (i.e. one red emitter in the right bottom corner) in the following runs. One can see, that the ramp down $I-E$ curve of the initial run (Cb) and $I-E$ curves of the subsequent run (Ca) are identical.

There was no quantification metric for emission area of the sample D and its emission imaging results will be discussed in more detail in subsection \ref{emission_uniformity} that follows.

Overall, the emission improved after fibers underwent conditioning breakdowns: this is seen as improved efficiency (lower turn-on field and enhanced $\beta$-factor) which happened alongside with the improved spatial emissivity of the fibers in that the emission area was increased. A rough stepwise process can be described as 1) the breakdown increases the number of emitters (i.e. through mechanical unfolding); 2) the larger number of emitters deliver larger output current seen as lowered turn-on field and larger $\beta$-factor of the FN like part of the $I-E$ curves. This model is further supported by the analysis of the saturation sections of the $I-E$ curves; saturation follows the FN-like part when going to higher current range. The FN-like and saturation regions are labeled with \texttt{FN} and \texttt{sat.} respectively on the FN plots given in Fig.\ref{f3}.

\subsection{Emission area: FN vs image processing}

According to the Fowler-Nordheim (FN) law, the emission current as a function of applied electric field is given by:

\begin{equation}\label{e1}
I=1.54\times10^{-6}\left(\frac{\delta S}{\phi}\right)(\beta\cdot E)^2\cdot exp\left(\frac{-6.83\times 10^9\cdot \phi^{3/2}}{\beta\cdot E}\right)
\end{equation}
where $\delta S$ is effective emission area, $\beta$ is unitless effective field enhancement factor, and $\phi$ is the work function which is assumed as 4.8 eV for all the CNT fiber geometries. When $ln(I/E^2)$ is plotted against $1/E$, the slope gives [$-6.83 \times 10^9\cdot \phi^{3/2}/\beta$].

\begin{figure*}
\includegraphics[scale=0.16]{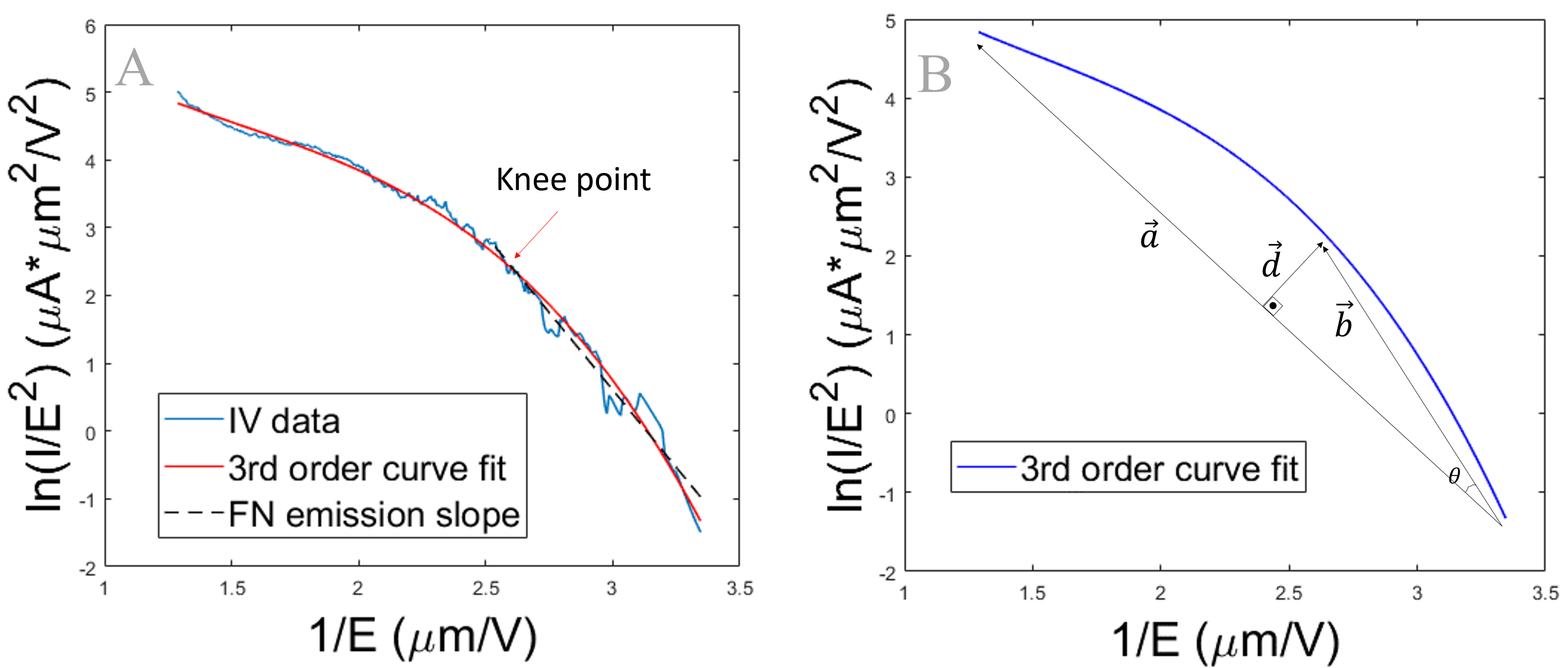}
\caption{\label{f5}a) Emission curve after noise reduction processing (blue), 3rd order polynomial fitting (red), deduced FN-like section of the $I-E$ curve used for calculation of $\mu$ factor (black); b) The vector family used in the knee point calculation.}
\end{figure*}

Although for metallic surfaces, the experimental data show linear slope\cite{13}, for the non-metallic and semi-metallic surfaces there is deviation of the slope from linear trend\cite{14}. For all geometries of the fiber samples tested, after filtering out conditioning noise portion of the FN plots, there were two distinct slope regions: one for low applied field and the other for high applied field (see FN plots in Fig.\ref{f3}). The curves have knee separating one slope region from another. The low applied field linear region of a higher slope corresponds to FN-like emission. The high applied field linear region of a lower slope corresponds to saturation region. The $\beta$-factors were calculated from the slope of FN-like portion of the $I-E$ curve using the following procedure: 1) noise data filtration, shown by blue solid line in Fig.\ref{f5}a; 2) third order polynomial fit, shown by red solid line in Fig.\ref{f5}a; 3) knee point calculation\cite{15}, using a set of vectors as shown in Fig.\ref{f5}b; 4) filtering out all the points above the knee point; 5) fitting fist order polynomial of the remaining low field curve and calculating its constant slope to extract effective field enhancement factor, the final slope is shown by black dashed line in Fig.\ref{f5}a. More specifically, step 3) when the knee point is calculated, is done by finding a unique point on the third order polynomial fit to find the maximum magnitude of vector$\overrightarrow{d}$ (shown in Fig.\ref{f5}b) defined as:
\begin{equation}\label{e2}
 \overrightarrow{d}=\overrightarrow{b}-b\cdot cos\theta \cdot  \hat{a}=\overrightarrow{b}-(\overrightarrow{b}\cdot \hat{a})\cdot \hat{a}
\end{equation}
where $\overrightarrow{b}$ is a constant vector between two edges of the curve, $\overrightarrow{a}$ is a variable vector from one edge of the curve to each data point, and $\theta$ is angle between $\overrightarrow{a}$ and $\overrightarrow{b}$. Final $\beta$-factor values extracted for all samples are labeled in Fig.\ref{f3}.

\begin{figure}
\includegraphics[scale=0.50]{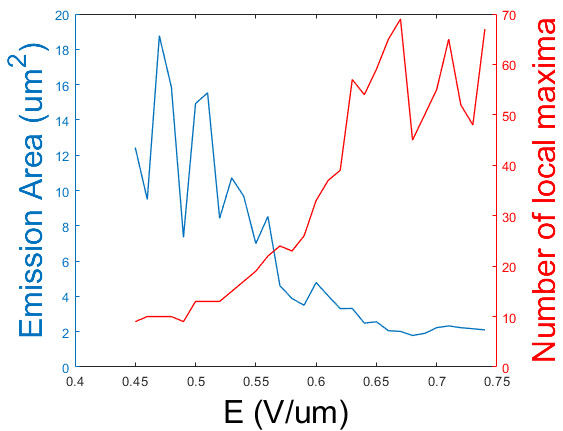}
\caption{\label{f6}Comparison of trends of the emission area on the applied electric field extracted from $I-E$ curves using FN equation versus from the field emission micrograph dataset using an image processing algorithm developed elsewhere.\cite{14}}
\end{figure}

Effective emission area is then calculated using the measured $I-E$ data and calculated $\beta$’s through Eq.\ref{e1}. The calculated dependence $\delta S(E)$ for sample B is shown in Fig.\ref{f6} as the decaying blue solid line. However, the results obtained using a custom image processing algorithm developed by our group before\cite{14} show opposite trend: $\delta S$ is predicted to increase as the applied field increases (red solid line in Fig.\ref{f6}). Field emission micrographs taken concurrently with $I-E$ curves and processed in batches point out that local emitting maxima multiply with the field. To keep the discussion to the general level and compare the trends, we do not present detailed analysis of the emission area and only calculate local maxima (brightest emitter locations). Assuming that the source behind every local maximum is a single CNT, $\delta S$ must grow with the field. The same issue was first pointed out in an original AFRL study of a CNT fiber\cite{1}. Using a PIC simulation, it was shown that the emission area has to be a growing function of E-field to account for the observed emission characteristics. The presented results are an experimental evidence that support the earlier PIC findings. This result is also supported by our earlier studies of nanodiamond emitters in which $\delta S$ grows nonlinearly with the electric field\cite{14}. Together, this result adds to concerns raised in recent literature\cite{16, 17} about the validity of FN equation application for extracting the emission area. This problem is under intense investigations in our lab.

\subsection{Current saturation}
One of the quantitative ways\cite{18} to describe saturation current plateau of a nonmetallic field emitter, or the total current limit that cannot be exceed, is:

\begin{equation}\label{e3}
    I_{s}^{max}=\frac{|e|\cdot n^{2/3}\cdot\upsilon_{\infty}}{l}\cdot\delta S
\end{equation}
where e is the electron charge, n is the bulk charge carrier concentration, $\upsilon_{\infty}$ is the saturated drift velocity, $l$ is the depletion length and $\delta S$ is the emission area. Since each sample fibers’ constituent CNT material is the same in the before and after experiments, it can be speculated that their properties are the same. The only parameter that is changing then in formula \ref{e3} is the emission area. From comparing the $I-E$ curves and field emission micrographs (given for before and after runs at the same electric field) for samples A and B, it is seen that the output current saturation plateau value increased by 5-7 times, and so did the emission area as suggested by image processing illustrated in the previous section. Unlike A and B, sample C showed no change of the output current in the saturation regime (ramp down of the Cb and Ca in Fig.\ref{f3}). From comparing to Fig.\ref{f4} Cb and Ca, it can be noted that the single emitter generated during breakdown remained and therefore no change in the output current was observed; this is consistent with formula \ref{e3}.

Using typical numbers for CNTs, $n\sim 10^{18}$ cm$^{-3}$ (calculated from $\sigma=e\cdot n\cdot \mu$ where $\sigma$=1 kS/cm \cite{19}, and $\mu$=10$^4$ cm$^2$/V$\cdot$s \cite{20}), $\upsilon_{\infty}\sim10^7$ cm/s \cite{21}, $l\sim890$ nm (calculated using Ref.\cite{22}), it yields the diameter of the emitter of 0.7 $\mu$m. This result is much smaller than the lateral resolution of our microscope yet detecting that single emitter as $\sim$0.1 mm spot on the YAG screen. The reason for that is a fairly large magnification of the system when the fibers are placed far away from the screen (1 mm in this case). Magnification of a point like electron source can be estimated as

\begin{equation}\label{e4}
    mag=2\cdot d\cdot tan(\alpha)
\end{equation}
where $d$ is the distance between the emitting surface and the screen ($d$ was 1 mm for samples A, B and C) and the angle $\alpha$ is calculated as

\begin{equation}\label{e5}
    \alpha=\frac{p_x}{p_z}=\sqrt{\frac{2\cdot MTE}{m_0\cdot c^2}}\cdot \frac{1}{\beta\cdot \gamma}
\end{equation}
where $p_{x}$ and $p_{z}$ are transverse and longitudinal momenta, $MTE$ is the mean transverse energy, $m_0c^2$ is the rest energy (0.511 MeV), $\beta$ is the ratio between electron velocity and the speed of light, and $\gamma$ is the Lorentz factor. By using $\beta=0.063$ and $\gamma=1$ at 1 kV and $MTE=4.5$ meV (corresponding to its Fermi energy), we find that a point like emitter would appear as a 0.25 mm spot on the screen due to spreading electron rays that have non-zero transverse momentum.

In Ref.\cite{1}, CNT fiber was found to saturate at about 250 $\mu$A. This result could not be explained by the vacuum space-charge (Child-Langmuir) effect. Extending the application of Eq.(\ref{e3}), we find that the emission area in saturation had to be $\sim$0.02\% of the total cross section area of the fiber. PIC results suggested $\sim$0.3\%. The order of magnitude discrepancy could be explained within the series ballast resistor model\cite{18,23}, by adding extra terms (in addition to the basic resistivities associated with the transport through the depletion region and tunneling barrier transparency) in the following form:

\begin{equation}\label{e6}
    R_i\propto \sum_{i}\frac{m^{\ast}}{e^2\cdot n}\cdot \frac{1}{\tau_i}
\end{equation}
where $\tau_i$ is a characteristic scattering time that should be associated with microscopic bundle/alignment structure. An effect of this sort, earlier observed in Ref.\cite{10}, can change the onset of saturation by many folds and adds an uncertainty to emission area calculation and was not included before into basic/simplified models.

\subsection{Emission uniformity and directionality}\label{emission_uniformity}
\begin{figure}
\includegraphics[scale=0.27]{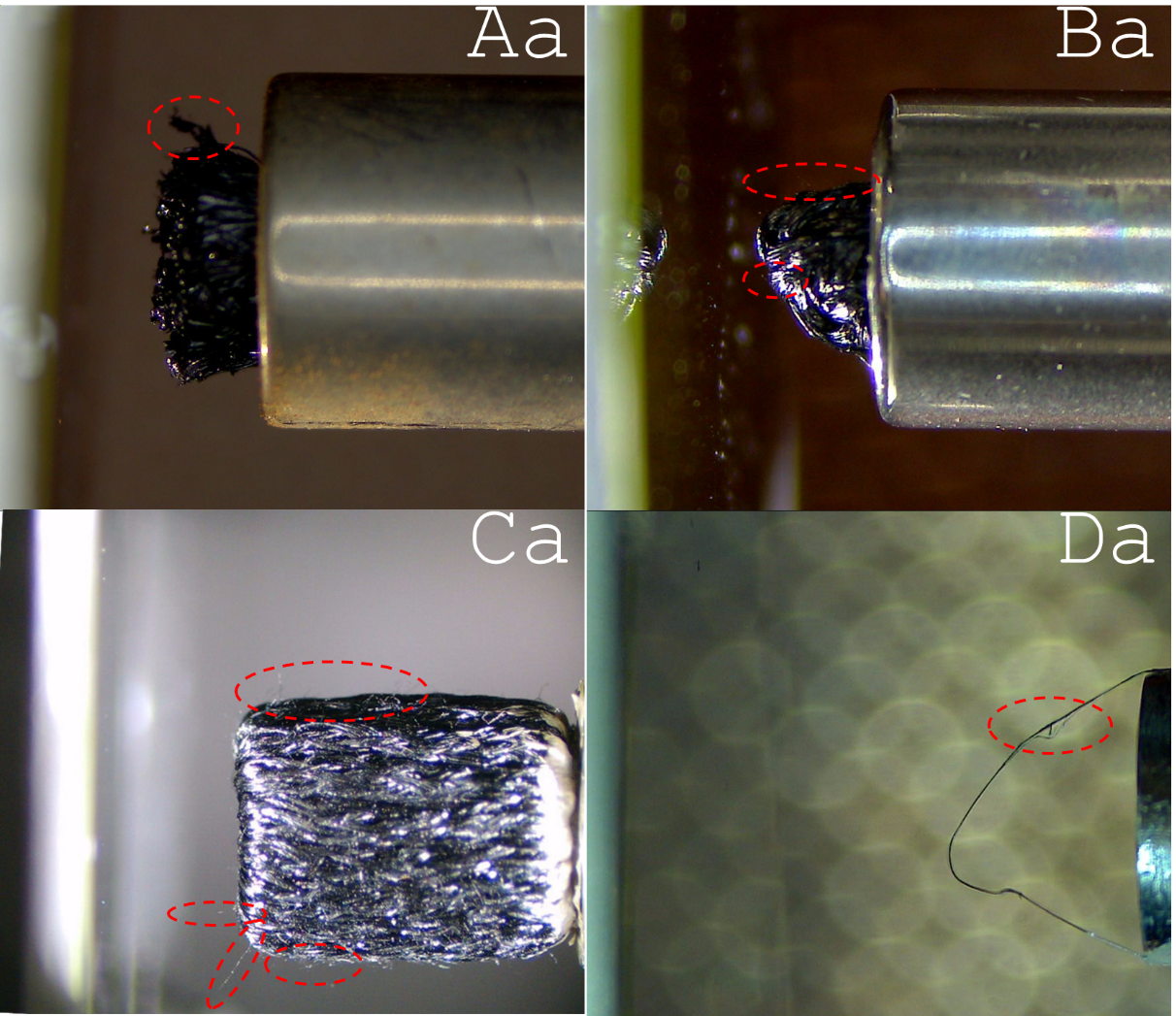}
\caption{\label{f7}Demonstration of various types of unfolded and differently aligned stray CNT fibrils that form after conditioning breakdowns – all taken by the top view camera.}
\end{figure}

Additional analysis of Fig.\ref{f4} shows that four fiber designs demonstrated very different emission patterns that can be further discussed and interpreted as follows:

1) There is the glowing background that exists on every image set in Fig.\ref{f4}. They come from tangent electron rays that penetrate the anode screen at shallow angles. Sometimes they can be visualized by moving the screen such that the sample is at the edge of the screen, or by increasing the distance between the screen and the cathode. Then the background halo source can be seen at the opposite edge of the screen as a streaked magnified nanotube oriented more in parallel, rather than perpendicular to the screen plane. This is exemplified in the outstanding image in Fig.\ref{f4} for sample C that had the strongest halo. Typically, the halo becomes stronger after the conditioning process, additionally confirming mechanical untangling of CNTs comprising the fibers. These CNT bunches are seen (highlighted by the dashed circles in Fig.\ref{f7}) by the top view camera measuring the interelectrode gap. One of them for sample C, bottom CNT bunch perpendicular to the screen, can be identified as the major emitter on the laterally resolved images in Fig.\ref{f4} – when image is taken by the top camera in dark, this location is glowing bright red corresponding to a black body temperature of 1500-2000 K. Looking at Fig.\ref{f4}, there is correlation: if the fiber is enclosed into a hollow cylinder like samples A and B, the background is suppressed suggesting mechanical support somewhat mitigates the untangling.

2) The physical dimensions are not directly related to emission properties, i.e. emission area is not necessarily large for a large size sample, such as exampled by sample C. All samples demonstrated a counted number of strong emitters during the “before” runs with emission area being orders of magnitude lower that the physical area available for emission.

3) Sample B, even improving the emission area upon conditioning, shows very large distribution of emission angles. When placed 1 mm away from the screen, the emission envelope is three times larger than the actual emitter size (dashed circle in Fig.\ref{f4} Ba). This suggests proper performance for MHz application unless the fiber is placed in a solenoid field for focusing. If X-band or beyond applications are sought, the brightness of such a design will deteriorate the performance of an VED.

4) Upon conditioning, sample D shows no stray emitters (Fig.\ref{f4}) and the emission pattern is an arch showing a coherent emission from a section of the looped fiber. At this point, there is no good procedure of evaluating exact field emission area for this type of geometry without knowing the emitting section. One main complication is the parallel shift with respect to the actual fiber loop position due to a fairly large emission angle. Compared to Fig.\ref{f4} Db, showing spatially incoherent emission centered near the actual emitter location, the emission pattern located away from the emitter (Fig.\ref{f4} Da) is caused by a mechanical bent when the front half section of the loop bent down. Even establishing coherent emission, this free-standing design shows weak resilience to conditioning breakdowns that always take place.

\begin{figure}
\centering
\includegraphics[scale=0.43]{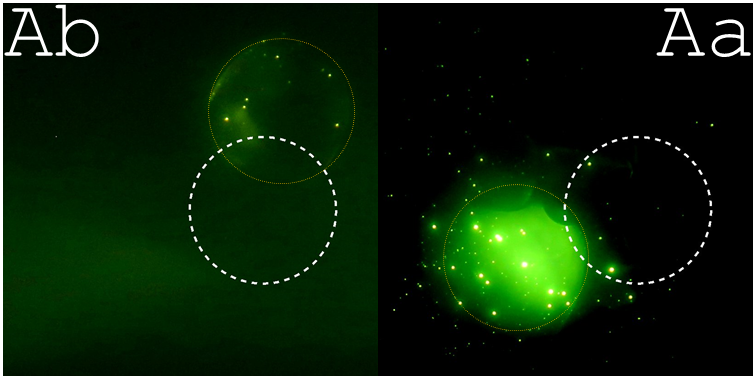}
\caption{\label{f8}Close-ups of emission patterns of sample A before and after the conditioning runs. The thicker dashed line circles depict the actual fiber location with respect to the YAG screen. Ultra-thin orange circles of the same diameter are to illustrate that major emission pattern fits within the size of the fiber even though there is a parallel shift caused by slight misalignment.}
\end{figure}

5) The most remarkable and promising performance was demonstrated by sample A. During the conditioning run, the emission is limited to $\sim$10 bright strongest point-like emitters combined with more distributed lobes. Most importantly – all the emission locations are confined within a circle of the size of the emitter A, see Fig.\ref{f8}. The parallel shift is due to cut and installation angle imperfections. Electrons start at a small angle but soon after, travel along uniform field lines yielding the projection shift with respect to the actual emitter location. In Fig.\ref{f8}, the white circle corresponds to the actual fiber position and the vanishingly thin orange circle of the same diameter is to emphasize that the diameter of the emission core matches the fiber diameter. After the breakdown, the parallel shift has changed due to the change in the relative position between the fiber and the screen. More bright point-like emitters appeared outside the emitter physical size boundary due to untangled stray emitters, but the core retained its shape and became brighter and more uniformly distributed carrying more emission current. Altogether, sample A demonstrated the best spatial emission coherence. Therefore, this design may be further optimized to achieve high brightness to be used as a driving injector for miniature/small size VEDs operated between X- and W- bands.

\section{Conclusion and outlook}

Field emission microscopy of four different CNT fiber designs is presented. Details of cathode conditioning upon the initial turn on are outlined. It is emphasized that the electrical breakdown plays critical role in establishing emission performance and operating point of the emitter, typically improving performance in terms of integral $I-E$ characteristic in that the turn-on field drops, field enhancement and emission area increases, the saturation level increases allowing for larger output current. The flat cut fiber geometry enclosed in a supporting tubing enclosure was found as a best design. Folded and wound designs either demonstrated lower spatial coherence or greatly suppressed area of emission due to unfolded stray CNT emitters after undergoing conditioning breakdowns, either would deteriorate performance when driving a high frequency VED. The free standing looped design showed weak mechanical stability against breakdown: while still promising additional design considerations must be made to strengthen its stability. Altogether, the new results support earlier findings and provide new insights into performance of the CNT fibers as the material-of-choice for future VED architectures/platforms.

\bibliography{taha_posos}

\end{document}